\documentclass{article}
\usepackage[T1]{fontenc}
\usepackage{spconf,amsmath,hyperref}
\usepackage{amsmath,amsfonts}
\usepackage{algorithmic}
\usepackage{graphicx}
\usepackage{textcomp}
\usepackage{xcolor}
\usepackage{hyperref}
\usepackage{cite}
\usepackage{multirow}
\usepackage[colorinlistoftodos]{todonotes}
\usepackage{hyperref}
\usepackage{tikz}
\usepackage{bm}

\usetikzlibrary{backgrounds,fit,positioning,calc}

\def\BibTeX{{\rm B\kern-.05em{\sc i\kern-.025em b}\kern-.08em
		T\kern-.1667em\lower.7ex\hbox{E}\kern-.125emX}}

\begin{document}
	\title{Targeted Pooled Latent-Space Steganalysis\\applied to Generative Steganography,\\ with a fix}
	\customauthors
	{Etienne Levecque$^{3}$, Aurelien Noirault$^{1}$, Tomas Pevny$^{2}$, Jan Butora$^{1}$, Patrick Bas$^{1}$, Rémi Cogranne$^{3}$}
	{$1.$ UMR 9189 CRIStAL \\
		Univ. Lille, CNRS, \\
		Centrale Lille, Lille, France }
	{$2.$ Artificial Intelligence Center \\
		Czech Technical University\\
		Prague, Czech Republic.}
	{$3.$ LIST3N, UTT\\
		University of Techn. of Troyes,\\ Troyes, France}

	\maketitle
	\begin{abstract}
		Steganographic schemes dedicated to generated images modify the seed vector in the latent space to embed a message. Whereas most steganalysis methods attempt to detect the embedding in the image space, this paper proposes to perform steganalysis in the latent space by modeling the statistical distribution of the norm of the latent vector. Specifically, we analyze the practical security of a scheme proposed by Hu {\it et al.}~\cite{hu2024establishing} for latent diffusion models, which is both robust and practically undetectable when steganalysis is performed on generated images. We show that after embedding, the Stego (latent) vector is distributed on a hypersphere while the Cover vector is i.i.d. Gaussian. By going from the image space to the latent space, we show that it is possible to model the norm of the vector in the latent space under the Cover or Stego hypothesis as Gaussian distributions with different variances. A Likelihood Ratio Test is then derived to perform pooled steganalysis. The impact of the potential knowledge of the prompt and the number of diffusion steps is also studied. Additionally, we show how, by randomly sampling the norm of the latent vector before generation, the initial Stego scheme becomes undetectable in the latent space.
	\end{abstract}
	
	\begin{keywords}
		Steganalysis, Generative Steganography, Latent space, statistical tests
	\end{keywords}

	\section{INTRODUCTION}
	
	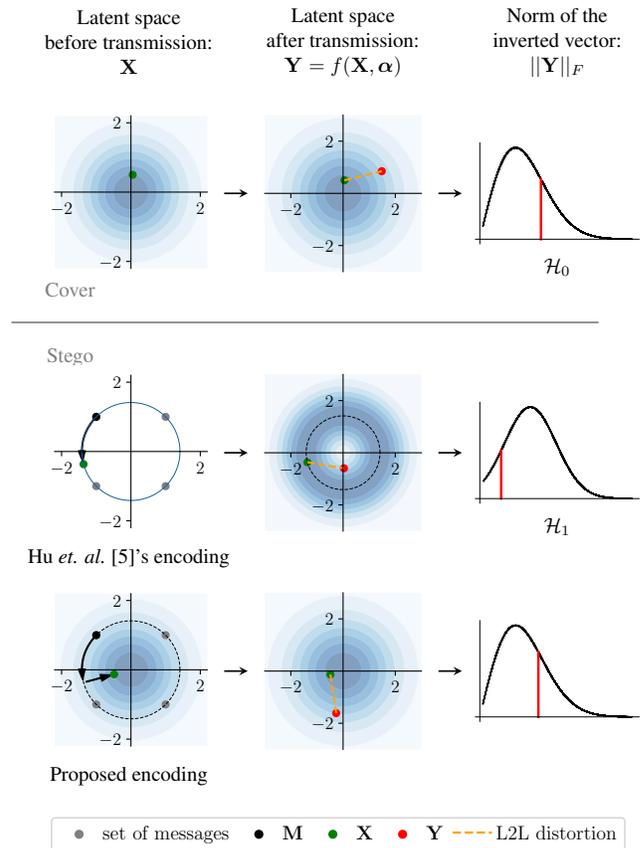
\begin{figure}[ht!]
		\centering
		\resizebox{\columnwidth}{!}{
			\begin{tikzpicture}[node distance=0.4cm]
				
				\tikzstyle{arrow} = [thick,->,>=stealth]
				
				\draw[gray, thick] (-2,-2.2) -- (8,-2.2) node[above, xshift=-9cm, yshift=0.3cm] {Cover} node[below, xshift=-9cm, yshift=-0.3cm] {Stego};
				
				\node (cover0) {\includegraphics[width=3cm]{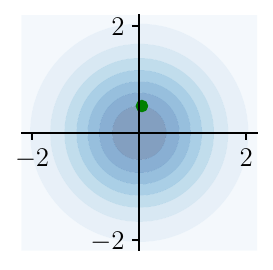}};
				\node[align=center, right=of cover0] (cover1) {\includegraphics[width=3cm]{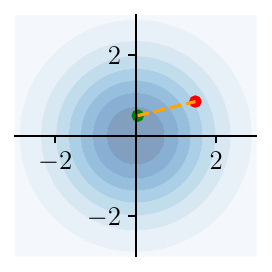}};
				\node[align=center, right=of cover1] (cover2) {\includegraphics[width=3cm]{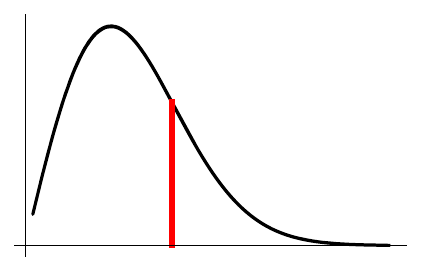}};
				\node[align=center, below=of cover2, yshift=0.5cm] (h0) {$\mathcal{H}_0$};
				
				\node[align=center, above=of cover0, yshift=-0.1cm] (latent0) {Latent space\\before transmission:\\$\mathbf{X}$};
				\node[align=center]  at (latent0 -| cover1) (latent1) {Latent space\\after transmission:\\$\mathbf{Y} = f(\mathbf{X}, \bm{\alpha})$};
				\node[align=center]  at (latent0 -| cover2) (norm) {Norm of the\\inverted vector:\\$||\mathbf{Y}||_F$};

				\node[align=center, below=of cover0, yshift=-0.8cm] (Stego1) {\includegraphics[width=3cm]{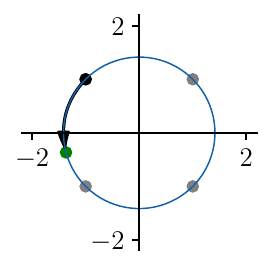}};
				\node[align=center, right=of Stego1] (Stego2) {\includegraphics[width=3cm]{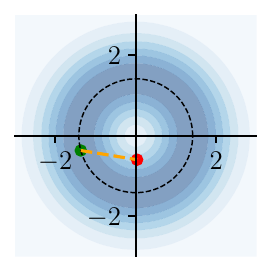}};
				\node[align=center, right=of Stego2] (Stego3) {\includegraphics[width=3cm]{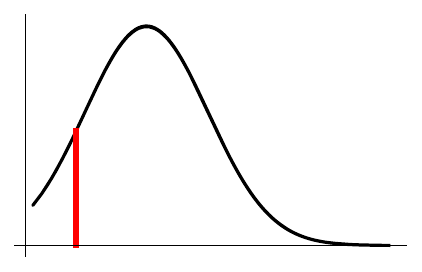}};
				\node[align=center, below=of Stego3, yshift=0.5cm] (h1) {$\mathcal{H}_1$};

				\node[align=center, below=of Stego1, yshift=-0.1cm] (Stegofix1) {\includegraphics[width=3cm]{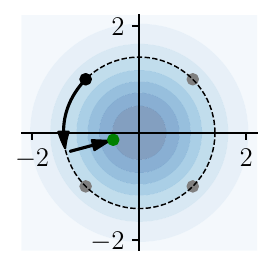}};
				\node[align=center, right=of Stegofix1] (Stegofix2) {\includegraphics[width=3cm]{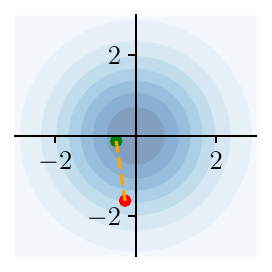}};
				\node[align=center, right=of Stegofix2] (Stegofix3) {\includegraphics[width=3cm]{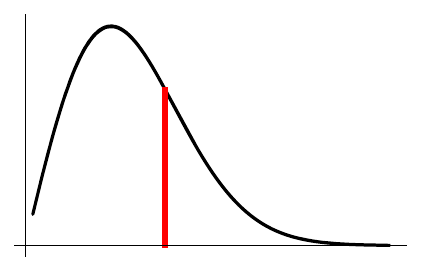}};
				
				\node[align=center, below=of Stegofix2] (legend) {\includegraphics[width=10.3cm]{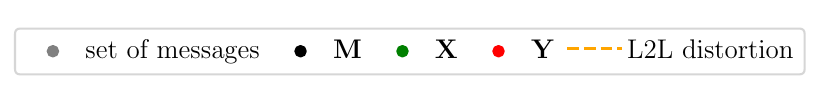}};
				
				\node[align=right, below=of Stego1, yshift=0.5cm](){Hu {\it et. al.}~\cite{hu2024establishing}'s encoding};
				\node[align=right, below=of Stegofix1, yshift=0.5cm](){Proposed encoding};
				
				\draw[arrow] (cover0.east) -- (cover1.west);
				\draw[arrow] (cover1.east) -- (cover2.west);
				
				\draw[arrow] (Stego1.east) -- (Stego2.west);
				\draw[arrow] (Stego2.east) -- (Stego3.west);
				
				\draw[arrow] (Stegofix1.east) -- (Stegofix2.west);
				\draw[arrow] (Stegofix2.east) -- (Stegofix3.west);
		\end{tikzpicture}}
		\caption{Illustration in 2D: gray dots in the Stego setup represent the space of messages using the encoding proposed in~\cite{hu2024establishing} (2 bits are encoded), which creates a spherical distribution in the original latent space and can be approximated as a Gaussian distribution after the latent-to-latent channel. Our proposal, consisting of sampling the norm after encoding, fixes this discrepancy.
		}
		\label{fig:dataset_flow}
		\vspace{-0.7cm}
	\end{figure}
	
	\begin{figure*}[t]
		\begin{centering}
			\includegraphics[width=\textwidth]{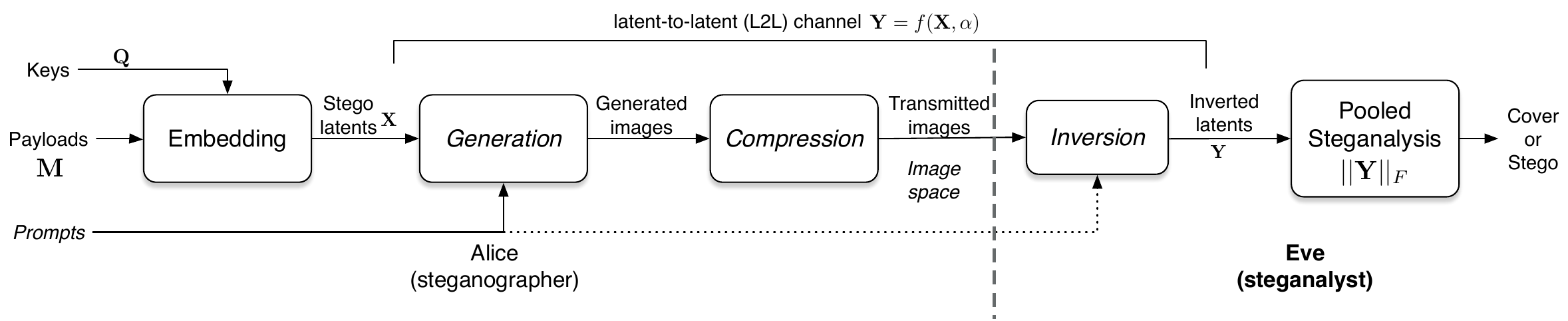}
			\vspace{-0.7cm}
			\caption{\label{fig:principle}Principle of the presented steganalysis scheme targeted to~\cite{hu2024establishing}. The entities in {\it italic} are, according to the Kerckhoffs principle, considered as public. Prompts are potentially public (dotted arrow).}
		\end{centering}
	\end{figure*}
	
	Steganography on generated images is a recent active field since it enables the embedding of large payloads and is rather robust to operations such as JPEG compression, while being possibly undetectable. One notable feature of generative steganography is the fact that, contrary to classical steganography~\cite{fridrich2009steganography}, which only alters the noise component of the image, the embedding also alters its semantic content. 
	Generative steganography methods~\cite{yang2024gaussian,zhou2025improved,hu2024establishing} are popular among latent diffusion models~\cite{rombach2022high} (LDM) since these models provide high-quality images. In this case, the embedding strategy consists of partitioning the latent space (typically of size $[4,64,64]$) into key-dependent and message-dependent regions and moving the seed into such a region before generation. Unfortunately, the whole latent-to-latent (L2L) process (consisting of generating, coding, and going back to the latent space) can be considered as a noisy process and may produce decoding errors. Consequently, one way to preserve robustness is to modify the distribution of the initially i.i.d. Gaussian distribution in a secret key-dependent subspace of the latent space in order to create clusters around the codewords related to each embedded message. 
	
	If this alteration may, in principle, increase the detectability of the Stego scheme, many published schemes are blind to security analysis in the latent space. The first class of steganalysis schemes assesses the detectability by training classical steganalysis networks~\cite{boroumand2018deep,Yed18} to distinguish between generated Cover\footnote{for the sake of simplicity and consistency with more than 20 years of steganography, we decide to keep naming non-Stego images as "Covers".} images and generated Stego images.
	The second class of security analysis proposes to compute the Kullback-Leibler Divergence on marginals in the latent space~\cite{zhang2019generative,zhang2019invisible} in order to show that the embedding has no impact on the distribution. 
	
	Our contribution focuses on the generative steganographic scheme proposed by Hu~{\it et. al.} which is both robust to the inversion channel (the Bit Error Rate is on average smaller than 2\%) and not detectable by steganalysis in the image domain. The originality of our detection scheme relies on the fact that the steganalysis is neither performed in the image space nor on marginal distributions in the latent space, but on the joint distribution in the latent space. Moreover, since the detection is based on a Likelihood Ratio Test (LRT), it is easy to derive a pooled detector based on a batch of images.
	\vspace{-0.3cm}

	\section{Binary mapping in the latent space }\label{sec:binary:mapping}
	
	\subsection{Notations and security hypothesis}
	
	Let $\mathbf{M}$ be a secret message of $n$ bits cast to $\pm 1$ and reshaped to match the latent space dimension $(4, \sqrt{n} /2, \sqrt{n} /2)$. In this paper, the latent space is of dimension $(4, 64, 64)$ and thus $n = 16384$.
	The latent-to-latent (L2L) channel is composed of three parts always in the same order: a generation process that maps the seed to an image, an operation inducing distortion such as compression, and the inverse generation that maps an image back to a vector $\mathbf{Y}$ in the latent space. This channel is noted $f$ such as $\mathbf{Y} = f(\mathbf{X}, \bm{\alpha})$ with $\bm{\alpha}$ a vector of hyperparameters such as the prompt, the guidance, or the compression.
	
	In this steganography setup, Alice and Bob have to agree on a secret key $\mathbf{k}$ and on a set of parameters $\bm{\alpha}$ which will be fixed for a given message. Note, however, that according to Kerckhoffs' principle, Eve, the steganalyst, \textit{has access} to the L2L channel $f$ and its fixed parameters $\bm{\alpha}$ but \textit{does not have access} to the secret key.
	
	\subsection{Description of the embedding scheme}
	
	This scheme, presented by Hu {\it et. al.}~\cite{hu2024establishing}, is particularly suited for LDM since it is robust to the L2L channel. It relies on the principle of Spread Spectrum watermarking~\cite{Cox07} (SS) where each embedded bit is spread on a secret carrier. To embed a secret message, Alice (the steganographer) performs the modulation using an orthonormal key-dependent pseudo-random matrix $\mathbf{Q}$ (it is the result of a Gram-Schmidt decomposition) using the following multiplication:
	\begin{equation}\label{eq:modulation}
		\mathbf{X} = \mathbf{Q}\cdot \mathbf{M} \cdot \mathbf{Q}^T,
	\end{equation} 
	Note that the matrix representation of the message $\mathbf{M}$ enables $\mathbf{Q}$ to have a separable transform where each carrier associated to the binary modulation $m_{i,j}\in \{-1,+1\}$ equals $\mathbf{C}_{i,j} = \mathbf{q}_i \cdot \mathbf{q}_j^T$, with $\mathbf{q}_k$ the $k^{th}$ column of $\mathbf{Q}$. 
	
	This latent seed $\mathbf{X}$ is sent through the L2L channel $f$:
	\begin{equation}
		\mathbf{Y} = f(\mathbf{X}, \bm{\alpha}).
	\end{equation}
	
	Bob, the receiver, can then estimate the payload $\hat{\mathbf{M}}$ by projecting on each carrier with the following operation: 
	\begin{equation}
		\mathbf{\hat{M}} = \mathbf{Q}^{T}\cdot \mathbf{Y} \cdot \mathbf{Q}.
	\end{equation}
	
	\subsection{Model of the norm before L2L}
	
	Let $R_{\mathbf{X}}$ be the random variable associated with the norm of the latent vector before the L2L channel. We recall that, under the null hypothesis (image is Cover) $\mathbf{X} \sim \mathcal{N}(\mathbf{0}, \mathbf{I}_{n})$ with $\mathbf{I}_n$ the identity matrix. The norm of a Gaussian distribution is a Chi distribution with $n$ degrees of freedom and thus still under the null hypothesis, $R_{\mathbf{X}} \sim \chi_n$. Finally, since the dimension of the latent vector $n$ is large, we can use the following limit:
	\begin{equation}
		\lim_{n \to +\infty} R_{\mathbf{X}} \xrightarrow[]{d} \mathcal{N}\left(\sqrt{n}, \frac{1}{2}\right).
	\end{equation}
	
	Under the alternative hypothesis (image is Stego), $\mathbf{X} = \mathbf{Q}\cdot \mathbf{M} \cdot \mathbf{Q}^T$ and thus we have that $R_{\mathbf{X}} = ||\mathbf{M}||_F$ because the norm is invariant by rotation. Moreover, we know that $\mathbf{M}$ is distributed on the vertices of the $n$-dimensional hypercube with edge length 2. Therefore, the distance between any vertex and the origin is always half the diagonal of the hypercube, which equals $\sqrt{n}$. Under the Stego hypothesis, the norm of the latent vector before the L2L channel is constant and is equal to $\sqrt{n}$ and $R_{\mathbf{X}} \sim \delta_{\sqrt{n}}$ with $\delta$ the Dirac distribution.
	\begin{equation}
		\left\{
		\begin{aligned}
			\mathcal{H}_0&: R_{\mathbf{X}} \sim \mathcal{N}\left(\sqrt{n}, \frac{1}{2}\right),\\
			\mathcal{H}_1&: R_{\mathbf{X}}\sim \delta_{\sqrt{n}}.
		\end{aligned}
		\right.
	\end{equation}
	
	\subsection{Model of the norm after L2L}
	
	We define $R_{\mathbf{Y}}$ as the random variable associated with the norm of the latent vector after the L2L channel. We recall that $\mathbf{Y} = f(\mathbf{X}, \bm{\alpha})$. To model the norm $R_{\mathbf{Y}}$ of the latent vector $\mathbf{Y}$, we assume that the L2L channel induces a distortion $E(\bm{\alpha})$ on the norm of $\mathbf{X}$ and that this distortion follows a Gaussian distribution: $\varepsilon(\bm{\alpha}) \sim \mathcal{N}(0, \sigma^2(\bm{\alpha}))$ and thus, $R_{\mathbf{Y}} = R_{\mathbf{X}} + E(\bm{\alpha})$.
	\begin{equation}
		\left\{
		\begin{aligned}
			\mathcal{H}_0&: R_{\mathbf{Y}} \sim \mathcal{N}\left(\sqrt{n}, \frac{1}{2} + \sigma^2(\bm{\alpha})\right),\\
			\mathcal{H}_1&: R_{\mathbf{Y}}\sim \mathcal{N}\left(\sqrt{n}, \sigma^2(\bm{\alpha})\right).
		\end{aligned}
		\right.
	\end{equation}
	We can see that both hypotheses only differ in the variance, and this difference should vanish if the L2L is very noisy. On the contrary, if the L2L is more and more accurate, the difference of variances will become maximal.
	
	\subsection{Scaled SS to fix distribution discrepancy}
	
	With classical SS encoding, we have seen that, under the Stego hypothesis, $\mathbf{X}$ is uniformly distributed on the hyper-sphere of radius $\sqrt{n}$. However, under the Cover hypothesis, $\mathbf{X}$ should follow a Gaussian distribution $\mathcal{N}(\mathbf{0}, \mathbf{I}_n)$ and its norm should follow a $\chi_{n}$-distribution. So, the solution to transform the uniform hyper-sphere distribution into a standard Gaussian one is to sample a norm $s$ according to the $\chi_{n}$-distribution and then to scale the norm of the latent vector as follows:
	\begin{equation}
		\mathbf{X} = \frac{s}{\sqrt{n}} \mathbf{Q}\cdot \mathbf{M} \cdot \mathbf{Q}^T.
	\end{equation}
	With this fix, we can verify that the norm of $\mathbf{X}$ is equal to $s$ and thus follows a $\chi_{n}$-distribution:
	\begin{equation}
		||\mathbf{X}||_F = \frac{s}{\sqrt{n}} \times ||\mathbf{M}||_F =s.
	\end{equation}
	To recover the payload $\mathbf{\hat{M}}$ from the latent vector $\mathbf{Y}$, Bob needs to apply the inverse rotation and observes the sign of each coefficient:
	\begin{equation}
		\mathbf{\hat{M}} = \text{sign}\left(\mathbf{Q}^{T}\cdot \mathbf{Y} \cdot \mathbf{Q}\right).
	\end{equation}
	
\begin{figure}
		\centering
		\includegraphics[width=0.9\columnwidth]{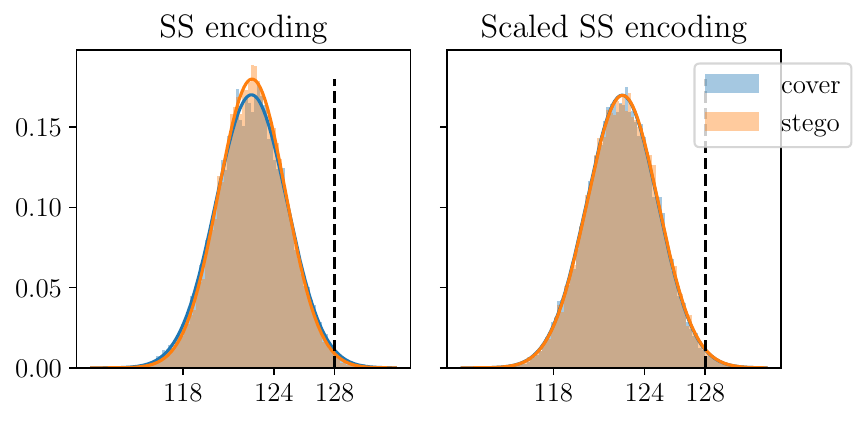}
		\vspace{-0.3cm}
		\caption{Histogram of the norm of the latent vector $||\mathbf{Y}||_F$. The curves represent the Gaussian fit for each class. Notice the small difference in the variance on the left plot, which is corrected on the right plot. Guidance=5.0, steps=20, 20k prompts.}
		\label{fig:histogram}
		\vspace{-0.3cm}
	\end{figure}

	Fig.~\ref{fig:histogram} illustrates how the fix acts on the norm of the latent vector $||\mathbf{Y}||_F$ received by Bob. Without the scaling, the Cover and the Stego distribution do not have the same variance as the norm, but with the scaling, this difference vanishes. Note that the mean is 122.5 for all 4 distributions instead of 128. With SS encoding, the estimated variance is 5.50 for the Covers and 4.91 for the Stegos, with a difference of 0.61. These small differences with the statistical model can be explained by the assumption on $f$.
	
	\vspace{-0.3cm}
	\section{Statistical steganalysis}\label{sec:results}
	\vspace{-0.2cm}
	
	\subsection{Likelihood Ratio Test}
	
	The proposed attack is based on a likelihood ratio test (LRT) on the norm $R_{\mathbf{Y}}$ of the latent vector. As shown previously, this norm follows a Gaussian distribution whose parameters depend on the hypothesis. In theory, the variance is the main source of discrepancy, while the mean should be identical. However, our theoretical model relies on assumptions about $f$, and as shown in Fig.~\ref{fig:histogram}, the mean is not exactly $\sqrt{n} = 128$ and the variance $\sigma^2(\bm{\alpha})$ is unknown. To address this, we estimate the mean and variance using Maximum Likelihood Estimation. Since the mean is assumed to be identical under both hypotheses, we denote by $\hat{\mu}$ the estimated mean of both the Cover and Stego distributions, and by $\hat{\sigma}^2_0$ (respectively $\hat{\sigma}^2_1$) the estimated variance of the Cover (respectively Stego) distribution. Let $r_{\mathbf{Y}}$ be a realization of $R_{\mathbf{Y}}$; the LRT for a single image is then given by:
	\begin{equation}
		\Lambda(r_{\mathbf{Y}}) = \frac{\mathcal{N}\left(r_{\mathbf{Y}}; \hat{\mu}, \hat{\sigma}^2_0\right)}{\mathcal{N}\left(r_{\mathbf{Y}}; \hat{\mu}, \hat{\sigma}^2_1\right)}.
	\end{equation}
	
	The null hypothesis is rejected if $\Lambda(r_{\mathbf{Y}}) < \tau$, where $\tau$ is a decision threshold, and the image is classified as Stego; otherwise, it is classified as Cover. The choice of $\tau$ is discussed in Section~\ref{sec:expe}.
	
	This LRT can be extended to a batch of images to increase its power. Assuming that a batch of $N$ images belongs to a single class (Cover or Stego), and denoting by $B = (r_{\mathbf{Y_i}})_i$ the batch of norms, the independence assumption yields the following LRT:
	\begin{equation}
		\Lambda(B) = \prod_{i=1}^N \frac{\mathcal{N}\left(r_{\mathbf{Y_i}}; \hat{\mu}, \hat{\sigma}^2_0\right)}{\mathcal{N}\left(r_{\mathbf{Y_i}}; \hat{\mu}, \hat{\sigma}^2_1\right)}.
	\end{equation}
	
	\vspace{-0.6cm}
	\subsection{Experimental results}\label{sec:expe}
	\begin{figure}[t]
		\begin{center}
			\includegraphics[width=0.9\columnwidth]{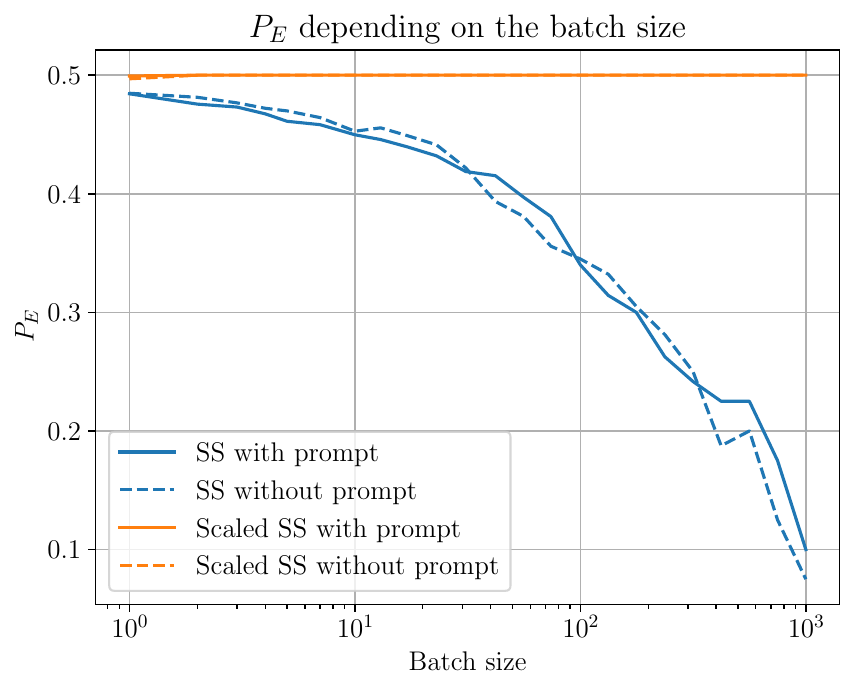}        
		\end{center}
		\vspace{-0.7cm}
		\caption{$P_E$ for steps = 20: as the batch size increases, the LRT becomes more powerful, contrary to the proposed fix, which is effective.}
		\label{fig:fix}
		\vspace{-0.4cm}
	\end{figure}
	
	All the experiments have been done with StableDiffusion 1.5 because it was the version used by Hu {\it et al.}, but the attack should work with any diffusion model. In their paper, they used a guidance of 5.0 for generation and inversion and 20 diffusion steps, so we fixed these parameters to the same values. The prompts are randomly selected from the DiffusionDB~\cite{wangDiffusionDBLargescalePrompt2022} database. To avoid any prompt mismatch, each randomly selected prompt was used to generate both a Cover image with a random seed $\mathbf{X}$ and a Stego image with a payload as the seed. A total of 20k prompts were used in each experiment, resulting in 20k Stego images and 20k Cover images.
	
	To evaluate our test, we used the Probability of Error ($P_E$) defined as:
	\begin{equation}
		P_E = \min_{P_{FA}} \frac{P_{FA} + P_{MD}(P_{FA})}{2},
	\end{equation}
	where $P_{FA}$ is the probability of false alarm and $P_{MD}$ is the probability of missed detection.
	
	Concerning the decision threshold $\tau$, the estimation of the mean and variance becomes noisy for larger batch sizes because fewer samples are available. This noisiness creates false alarms, so we scale the threshold based on the batch size. The expression of the threshold $\tau$ is:
	\begin{equation}
		\tau = \exp\left(-\log(N) / 20\right),
	\end{equation}
	with $N$ the batch size. If the estimation is not noisy, it is possible to keep $\tau = 1$ for any batch size. Finally, to evaluate the robustness of the Stego encoding, we use the Bit Error Rate ($BER$), defined as the proportion of correct bits decoded.
	
	For both experiments, we used a 20-fold cross-validation and averaged the $P_E$ across each fold. The estimated mean and variance of the norm are computed on 1k Cover and 1k Stego images, while the remaining 19k Cover and 19k Stego images are used to compute $P_E$. Fig.~\ref{fig:fix} shows the probability of error of the attack, which starts around 48\% for a single image and decreases with increasing batch size, while the proposed scaled SS encoding achieves a near-perfect 50\% $P_E$. Table~\ref{tab:steps} reports $P_E$ values for different numbers of diffusion steps for both generation and inversion in the L2L channel, showing that more diffusion steps reduce channel distortion and improve attack effectiveness. Moreover, as shown in Table~\ref{tab:ber}, the proposed scaled encoding does not alter the $BER$ of the original encoding. Finally, Fig.~\ref{fig:fix} suggests that even without knowledge of the prompt or the L2L parameters $\bm{\alpha}$, Eve can still perform the attack with the same effectiveness.
	
	\begin{table}
		\centering
		\renewcommand{\arraystretch}{1.1}
		\begin{tabular}{c|c|c}
			$P_E$ (\%)& \multicolumn{2}{c}{Steps}  \\ 
			\hline
			Batch size & 20   & 50                  \\ 
			\hline
			1          & 48.4 & 46.4                \\
			10         & 45.0 & 40.9                \\
			100        & 34.0 & 30.0               
		\end{tabular}
		\vspace{-0.2cm}
		\caption{$P_E$ in percentage for different numbers of diffusion steps.}
		\label{tab:steps}
		\vspace{-0.3cm}
	\end{table}
	\begin{table}
		\centering
		\renewcommand{\arraystretch}{1.1}
		\begin{center}
			\begin{tabular}{c|c|c}
				$BER$ (\%)& With prompt & Without prompt \\ \hline
				SS~\cite{hu2024establishing} & 97.53 & 97.56 \\ \hline
				Scaled SS & 97.53 & 97.52
			\end{tabular}
		\end{center}
		\vspace{-0.5cm}
		\caption{Impact of the knowledge of the prompt.
			Guidance=5.0, steps=20.}
		\label{tab:ber}
		\vspace{-0.5cm}
	\end{table}
	
	\vspace{-0.3cm}
	\section{Conclusions and perspectives}\label{sec:conclusions}
	\vspace{-0.2cm}
	
	This paper shows that if the message embedding for generative steganography is not detectable in the image space or on marginal distributions in the latent space, the inversion process makes the steganalysis still possible when relying on multivariate distributions, for example, by deriving a test on the norm of the vector, like in this contribution. 
	We also show that mimicking the multivariate distribution in the latent space, here by sampling the norm of the seed, is a way to remove this weakness and to achieve Stego-security~\cite{cayre2008kerckhoffs}, which means that the distributions of Cover and Stego vectors are identical.
	
	However, since the inversion channel is noisy, the detection process depends on inversion parameters, such as the number of steps or prior knowledge of the prompt. Given that the quality of the generative process heavily relies on the denoising processes, attacks in latent space will likely become more popular in the future.
	
	\pagebreak
	\section{Acknowledgments}
	This work received funding from the French Defense \& Innovation Agency. This work was also supported by a French government grant managed by the Agence Nationale de la Recherche under the France 2030 program, reference ANR-22-PECY-0011. This work was also supported by a grant managed by the Grant Agency of the Czech Republic, reference GACR 25-17259K.
	
	\bibliographystyle{plain}
	\bibliography{biblio.bib}

\end{document}